\documentclass[10pt,twocolumn]{article}

\DeclareUnicodeCharacter{3B1}{\ensuremath{\alpha}}

\usepackage{pdflscape} 

\usepackage{setspace}

\usepackage{stfloats}
\usepackage{float}
\usepackage{cuted}

\usepackage[percent]{overpic}  

\usepackage{setspace}

\usepackage{multicol,lipsum,microtype} 

\usepackage{authblk}  

\usepackage[T1]{fontenc}
\usepackage{palatino}
\usepackage{bold-extra}
\usepackage{bm}

\usepackage{amsmath}
\usepackage{graphicx}
\usepackage{booktabs}  
\usepackage{caption}
\usepackage{threeparttable}  
\usepackage{siunitx}  
\usepackage[table]{xcolor}  
\usepackage{setspace}  
\usepackage{cellspace}  
\usepackage{color}
\usepackage{colortbl}
\usepackage{textcomp}
\usepackage{siunitx}
\usepackage{threeparttable}
\usepackage{helvet} 
\usepackage{tikz}
\usepackage{threeparttable}
\usepackage{bigstrut} 

\definecolor{natcommtab1}{RGB}{214,214,194}
\definecolor{natcommtab2}{RGB}{236,235,226}
\definecolor{natcommtab3}{RGB}{251,250,246}

\usepackage{blindtext}

\usepackage{graphicx} 
\graphicspath{{figures}} 
\usepackage[font=small,labelfont=bf]{caption} 

\usepackage{xcolor}
\definecolor{black}{HTML}{212427}
\definecolor{blue}{HTML}{0563C1}
\definecolor{brightred}{HTML}{FF0000} 

\usepackage{hyperref,nameref}
\hypersetup{colorlinks=true,allcolors=blue}
\newcommand{\rref}[2]{\hyperref[#1]{\ref{#1}#2}} 

\color{black}
\usepackage[margin=0.67in]{geometry} 
\renewcommand{\thefootnote}{\fnsymbol{footnote}} 
\usepackage[switch]{lineno} 

\usepackage{fancyhdr,lastpage} 
\renewcommand{\headrulewidth}{0pt} 

\usepackage{titlesec}
\titlespacing{\section}{0pt}{10pt}{0pt}
\titlespacing*{\section}{0pt}{\baselineskip}{5pt}
\titlespacing{\subsection}{0pt}{10pt}{0pt}

\usepackage{amsmath,amssymb}
\renewcommand{\v}[1]{\boldsymbol{\mathbf{#1}}} 
\newcommand{\Ang}[0]{\mathring{\mathrm{A}}} 
\renewcommand{\l}[0]{\left} 
\renewcommand{\r}[0]{\right} 
\let\f=\frac 
\renewcommand{\t}[1]{\text{#1}} 

\usepackage{multirow} 
\renewcommand{\arraystretch}{1.3} 

\usepackage[defernumbers=true,style=nature,url=false,isbn=false,eprint=false]{biblatex}

\usepackage[normalem]{ulem}

\begin{document}

\twocolumn[
\begin{flushleft}
\fontsize{28}{33}\selectfont
\textbf{To crack, or not to crack: How hydrogen favors crack propagation in iron at the atomic scale}
\end{flushleft}
\vspace{0.35cm}
\bigskip 
\rule{\textwidth}{0.5pt} 

  Aleksei Egorov\textsuperscript{1},
  Lei Zhang\textsuperscript{2}, Erik van der Giessen\textsuperscript{1}, and
  Francesco~Maresca\textsuperscript{2,$\star$}\\
  \textit{\small \textsuperscript{1}Zernike Institute for Advanced Materials, Faculty of Science and Engineering, University of Groningen, Nijenborgh 3, 9747 AG, Groningen, The Netherlands \\ \small \textsuperscript{2}Engineering and Technology Institute Groningen (ENTEG), Faculty of Science and Engineering, University of Groningen, Nijenborgh 4, 9747 AG, Groningen, the Netherlands} \\ \small
  \textsuperscript{$\star$}Corresponding author: \url{f.maresca@rug.nl} (F. Maresca)

  \begin{center}
  \end{center}
  \vspace{-1.25cm}

\hfill
\rule{\textwidth}{0.5pt} 


\large
Steel is a key structural material because of its considerable strength and ductility. However, when exposed to hydrogen, it is prone to embrittlement. Mechanistic understanding of the origin of hydrogen embrittlement is hampered by the lack of reliable interatomic potentials. Here, we perform large-scale molecular dynamics simulations of crack propagation after having developed and validated an efficient yet density-functional-theory-accurate machine-learning potential for hydrogen in iron. Simulations based on our potential reveal that in the absence of H, iron is intrinsically ductile at finite temperatures with crack-tip blunting assisted by dislocation emission. By contrast, minute (part-per-million) hydrogen concentrations can switch the crack-tip behavior from ductile blunting to brittle propagation. Detailed analysis of our molecular dynamics results reveals that the combination of fast hydrogen diffusion and diminished surface energy is at the origin of embrittlement. Our results set the stage for a modified Griffith's criterion for hydrogen-induced brittle fracture, which closely captures the simulations and that can be used to assess embrittlement in iron-based alloys. 
\vspace{-0.05cm}
]

\begin{strip}
\rule{\textwidth}{0.5pt}
\vspace{-0.9cm}
\end{strip}


\begin{refsection}

\section*{\normalsize{Introduction}}
Hydrogen embrittlement (HE) has been known for a long time and is currently a critical bottleneck for the implementation of the hydrogen economy\textsuperscript{\cite{doi:10.1021/acssuschemeng.4c04328}
}. Notably, only a few ppm hydrogen in metals can lead to severe embrittlement\textsuperscript{\cite{wang2009effects_toughness_vs_h_content,dozen_times_more_brittle_LI201815575}}, potentially causing catastrophic failure. Hence, it is crucial to understand the key mechanisms giving rise to this phenomenon in order to mitigate it. Since the discovery of HE in 1875\textsuperscript{\cite{johnson1875ii}}, several hypotheses have been proposed for its mechanistic origin, yet there is no consensus\textsuperscript{\cite{pfeil1926effect_hede, westlake_hydrides_osti_4173745,Beachem1972, LYNCH1979_aide, BIRNBAUM1994191_help, vacancies_theory_2001590}
}. Amongst the proposed mechanisms of embrittlement, hydrogen-enhanced decohesion (HEDE) assumes that crack nucleation and propagation are favored by reduced surface energy due to hydrogen\textsuperscript{\cite{pfeil1926effect_hede,jiang_carter_ActaMat_2004_compute_surf_ener_with_H}}. However, the atomistic process that can lead to the enhanced decohesion has not been clarified to date.

{\it In situ} experimental validation of the proposed atomistic mechanisms is a daunting task, because hydrogen is hard to detect\textsuperscript{\cite{doi:10.1126/sciadv.aay4312}
} and, in body-centered-cubic (bcc) metals, diffuses rapidly\textsuperscript{\cite{Kim2025}}. 
Moreover, the reliability of atomistic simulations used to validate HE mechanisms is limited by the interatomic potentials, especially for the case of bcc metals like iron (Fe). 
In particular, empirical potentials are not capable of reproducing the correct structure and mobility of the defects that are most relevant to embrittlement, i.e. cracks\textsuperscript{\cite{moller_bitzek_2014comparative_eam_drawbacks,class_pot_limitations_Bitzek2015,eam_artifacts_HIREMATH2022111283,zhang2023atomistic}} and dislocations\textsuperscript{\cite{RODNEY20141591}
}.

For example, molecular dynamics simulations based on empirical potentials have shown that dislocation emission can be more favorable than cleavage in the presence of hydrogen, and that the mechanism responsible for embrittlement is hydride formation which inhibits the emission process\textsuperscript{\cite{Song_Curtin_NatMat_2013}}. Nevertheless, unlike e.g. titanium\textsuperscript{\cite{Kooi2020}}, hydrogen is not a hydride former and, to the best of the authors' knowledge, no {\it experimental} evidence of hydride formation causing embrittlement in iron has been reported. In fact, some atomistic simulations of hydride formation in iron feature a phase transition ahead of the crack tip, which is an artifact of the used interatomic potential\textsuperscript{\cite{moller_bitzek_2014comparative_eam_drawbacks,zhang2023atomistic}}. An overview of previous studies on atomistic crack simulations in hydrogen-charged iron is reported in the Supp. Info. S1 ``Literature on atomistic modelling of cracks with hydrogen''.

\begin{figure*}[htb!!]
  \centering
  \includegraphics[width=0.9999999\textwidth]{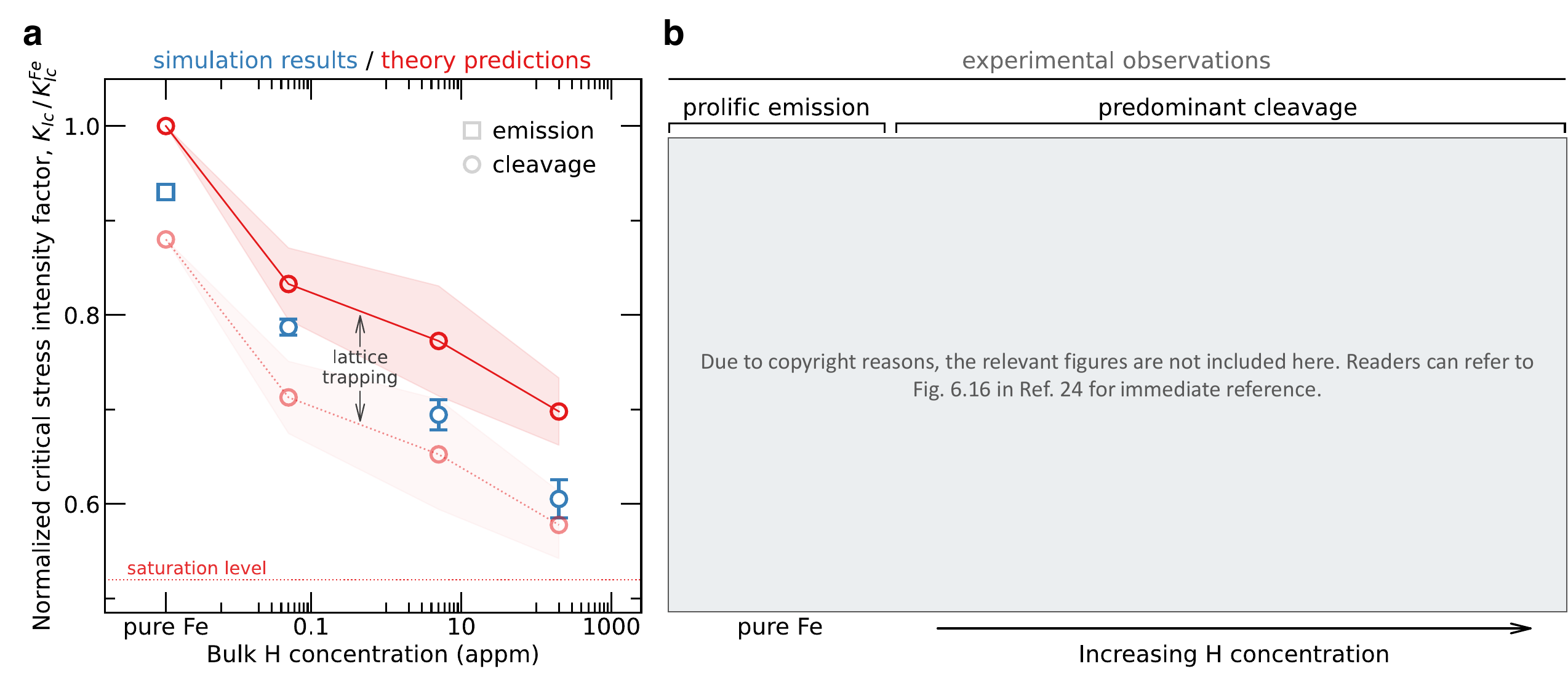}
  \caption{\label{proposed_theory} {\textbf{Hydrogen-induced crack propagation.} \textbf{a,} Critical stress intensity factor as a function of the bulk hydrogen concentration. MD simulation results obtained using the FeH ACE potential are compared with predictions from the proposed embrittlement theory, Eq.~\ref{Eq:theory}. The theory predictions are based on local hydrogen content ahead of the crack tip extracted from MD simulations (Fig.~\ref{bonds_and_h_atoms}), which determines the ``fresh'' crack surface coverage. The {\it saturation level} is the theoretical lower bound for a fully H-saturated fresh surface. Theory predictions (bright red circles and solid lines) include pure iron lattice trapping (second term in Eq.~\ref{Eq:theory}). As hydrogen weakens the iron-iron bonds, the lattice trapping becomes weaker as well\textsuperscript{\cite{TEHRANCHI_curtin_LT_2017150}}. Evaluating the impact of hydrogen on its strength is beyond the scope of this study, however, theory predictions without lattice trapping are shown as well and represent a lower bound (light red circles and dotted line). Indeed, MD simulation results lie between theory predictions with and without the (pure iron) lattice trapping. Notably, as hydrogen concentration increases, MD results approach theory predictions without lattice trapping. All results are presented as mean values of at least three simulations with 95\% confidence intervals. For the simulation results of pure iron, the confidence interval is smaller than the marker size. \textbf{b,} Experimental observations of the crack tip under mode-I loading, at varying hydrogen concentrations. Pure iron exhibits extensive dislocation emission at the crack tip. Note that dislocation traces appear as lines emanating at an angle from the crack surfaces. Increasing hydrogen concentration (hydrogen pressures of 0.7, 10, and 100 Pa) induces a transition from emission to predominant cleavage. Figure 1b refers to Fig. 6.16 in ref.~\cite{Vehoff1997}.}}\vspace{0.2cm}\end{figure*}

Here, we overcome the limitations of empirical potentials by developing a density-functional-theory (DFT)-accurate and efficient machine-learning (ML) potential for iron containing hydrogen (FeH). Inspired by experimental fracture tests of low-silicon, iron-alloy single crystals\textsuperscript{\cite{VEHOFF1980,VEHOFF1986_2nd}} (Figs.~\ref{proposed_theory}b), we examine hydrogen embrittlement of iron through atomistic simulations of the crack tip at varying bulk hydrogen concentration. In contrast with previous work based on empirical potentials and featuring hydride formation\textsuperscript{\cite{Song_Curtin_NatMat_2013}}, we observe that increased hydrogen concentration facilitates cleavage by lowering the critical stress intensity factor. This trend can be explained by a simple theory of embrittlement, whereby hydrogen locally lowers the surface energy at the crack tip due to fast hydrogen diffusion, facilitating cleavage (Fig.~\ref{proposed_theory}a).

\section*{\normalsize{Results}}

\subsection*{\normalsize{\textit{Interatomic potential development and validation}}}
\begin{figure*}[htb!!]
  \centering
  \includegraphics[width=0.69\textwidth]{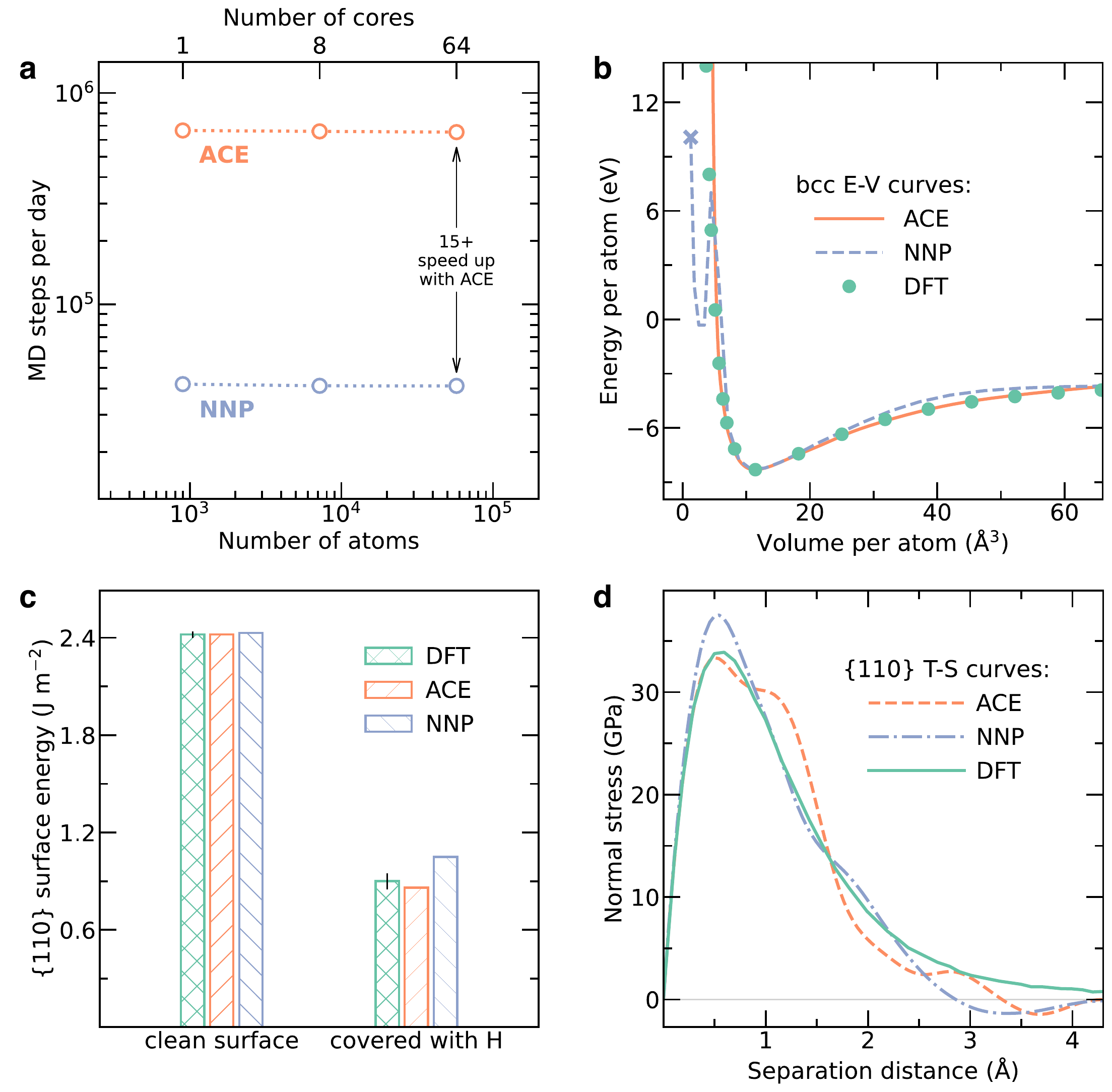}
  \caption{\label{basic_tests_fe} {\textbf{Performance of Fe-H Atomic Cluster Expansion (ACE) potential.} \textbf{a,} The ACE computational speed compared to neural network potential\textsuperscript{\cite{NNP_PRM_Ogata_2021}} (NNP), as MD steps per day as a function of the number of atoms and cores. The corresponding computational times (expressed in microseconds per timestep per atom) are 0.15, 0.15, and 0.14 for ACE; and 2.34, 2.34, and 2.29 for NNP. \textbf{b, c, d,} Primary validation tests against density functional theory (DFT) and NNP: \textbf{b,} The bcc energy versus volume (E-V) curves; the NNP produces the E-V curve with a spurious local minimum at small volumes, and it fails to converge at even smaller volumes, resulting in an error (cross). \textbf{c,} Energies of the H-free (left) and fully H-saturated (right) $\text{\{}110\text{\}}$ surfaces. DFT clean surface energies are shown as average values from refs.~\cite{surf_ener_1_JIN2022110029} and \cite{jiang_carter_ActaMat_2004_compute_surf_ener_with_H}; energies of the covered surfaces were computed following the procedure described in ref.~\cite{jiang_carter_ActaMat_2004_compute_surf_ener_with_H}, see Supp. Info. S3 ``Validation of the ACE FeH potential against benchmark properties''. DFT surface energies, both clean and covered, are reported with the corresponding 95\% confidence intervals. \textbf{d,} Traction--separation (T-S) curves for $\text{\{}110\text{\}}$ crack surfaces; notably, ACE captures the peak height, which limits the maximum stress that the material can withstand. More validation tests are in Supp. Info. S3 ``Validation of the ACE FeH potential against benchmark properties''.}}
\end{figure*}

To examine hydrogen embrittlement of iron via atomistic simulations, we first develop an atomic cluster expansion (ACE) interatomic potential. The ACE model is trained on a DFT database for the FeH system\textsuperscript{\cite{NNP_PRM_Ogata_2021}} (Supp. Fig. S1), which is extended here to achieve accurate energy--volume and traction--separation (T-S) curves, see Fig. \ref{basic_tests_fe}b,d, Supp. Info. S2 ``ACE FeH potential database and fitting'', Supp. Table S1 and Supp. Fig. S2. The ACE potential overcomes limitations of a previously developed neural network potential (NNP)\textsuperscript{\cite{NNP_PRM_Ogata_2021}}, which overestimates the cohesive strength and yields an unphysical local minimum along the energy--volume curve, see Figs. \ref{basic_tests_fe}b,d. Furthermore, the ACE potential has comparable accuracy as the NNP potential for other crucial properties including (i) hydrogen-dependent surface energies for $\{100\}$ and $\{110\}$ planes; (ii) hydrogen diffusion barriers; and (iii) hydrogen solution energies (Fig. \ref{basic_tests_fe}c and Supp. Info. S3 ``Validation of the ACE FeH potential against benchmark properties''). Finally, the ACE potential achieves $\sim15\times$ computational speed-up compared to NNP (Fig. \ref{basic_tests_fe}a): this gain paves the way for near-DFT-accurate and nanosecond-long simulations with hundreds of thousands of atoms, that are needed to simulate fracture at the atomic scale\textsuperscript{\cite{Andric_2019}}.

In particular, to be used confidently for crack-tip simulations, it is crucial that the ACE potential accurately reproduces several features of the T-S curve and hydrogen-dependent surface energy. Regarding the T-S curve, ACE matches DFT for key characteristics\textsuperscript{\cite{curtin_ts_curve_explained_2005methods}}, such as (i) the area under the T-S curve, which is the work of cohesion; (ii) the T-S peak, which is the cohesive strength of the material; and (iii) the maximum range of (nonlinear) interactions between the separating surfaces (Fig.~\ref{basic_tests_fe}d). Concerning the surface energy, ACE matches DFT predictions of both pure Fe and H-covered surfaces (Fig.~\ref{basic_tests_fe}c). Most notably, hydrogen is found to drastically lower the surface energy, which is computed using the approach devised in ref.~\cite{jiang_carter_ActaMat_2004_compute_surf_ener_with_H} (see Supp. Info. S3 ``Validation of the ACE FeH potential against benchmark properties''). A lower surface energy may favor crack growth over dislocation emission, yet hydrogen might also help in the emission of dislocations and thus suppress crack growth\textsuperscript{\cite{CLUM197551,Lynch+2019+reply}}. Therefore, direct atomistic simulations of near crack-tip processes are needed to clarify the competition between cleavage and blunting, in presence of hydrogen.

\subsection*{\normalsize{\textit{Atomistic fracture modeling}}}

\begin{figure*}[t!!]
  \centering
  \includegraphics[width=0.999\textwidth]{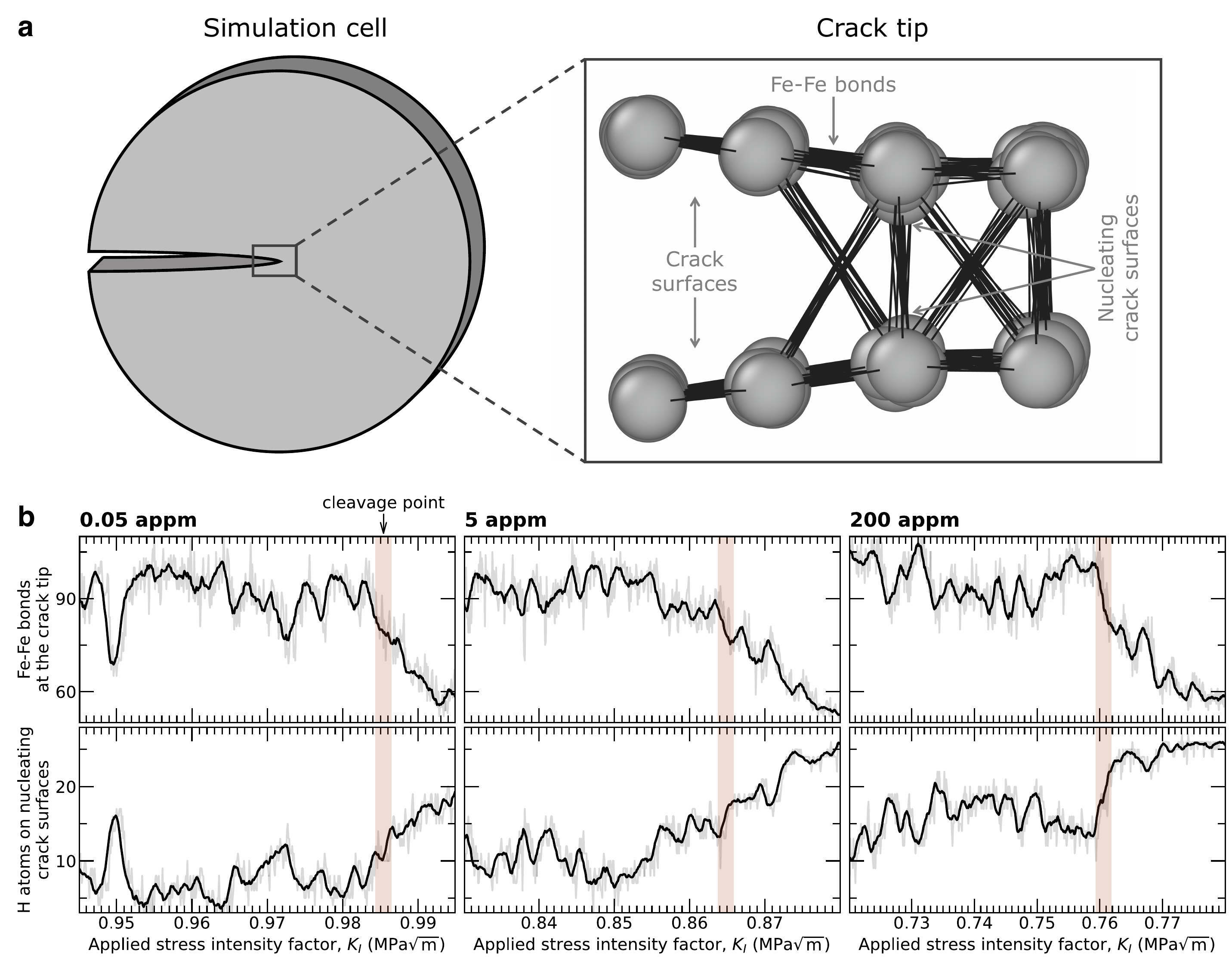}
  \caption{\label{bonds_and_h_atoms} {\textbf{The number of iron-iron chemical bonds at the crack tip and the number of hydrogen atoms on the nucleating crack surfaces.} \textbf{a}, A schematic illustrating the crack tip geometry and an inset with the atoms (highlighted in grey) ahead of the crack tip. \textbf{b}, Results of the MD simulations at various hydrogen bulk concentrations (0.05, 5, and 200 appm), showing the number of Fe-Fe bonds (top) and of H atoms approaching the newly forming crack surface, as a function of the increasing applied stress intensity factor. For each concentration, simulations were repeated at least trice, Here, representative cases are shown (see Supp. Info. S6 ``Hydrogen atoms count at crack tip and its relation to surface energy'' for the remaining cases). Light gray lines represent the actual data, while the thick black lines indicate the moving averages (see Supp. Info. S6 ``Hydrogen atoms count at crack tip and its relation to surface energy''). Cleavage points are marked where there is a sharp drop in the number of iron-iron bonds and are identified automatically by convolution with a step function\textsuperscript{\cite{convolution_smith1997scientist}}. We employed the \texttt{OVITO} package to count hydrogen atoms and bonds, setting a cutoff distance of 3.25~$\Ang$ for the largest bond (Fig.~\ref{basic_tests_fe}d); 
  See the details of counting bonds and hydrogen atoms in Supp. Info. S6 ``Hydrogen atoms count at crack tip and its relation to surface energy''.
}}
\end{figure*}

Equipped with a fast and robust ACE potential, we proceed with a simulation cell containing a sharp crack subject to mode-I loading imposed through boundary conditions\textsuperscript{\cite{Andric_2019}} (see Supp. Info. S4 ``Crack tip simulation setup'', Supp. Fig. S3 and Fig. \ref{bonds_and_h_atoms}a). A range of hydrogen concentrations is considered and the crack-tip mechanism is observed. We explore the (110)[110] crack system, where ($\cdot$) is the crack plane and [$\cdot$] the crack front, since this is often tested in experiments\textsuperscript{\cite{Nakasato1978_110_crack_surf,VEHOFF1986_2nd,Barnoush_WAN201987}}. We set the simulation temperature to $T=500~K$ because embrittlement still occurs\textsuperscript{\cite{stress_strian_curve_at_diff_temp}}, yet the quantum fluctuations that otherwise impact the behavior of light hydrogen in iron by increasing its diffusivity have faded away\textsuperscript{\cite{Cheng_Paxton_quntum_effects_PhysRevLett.120.225901}}. We control the applied load by increasing the mode-I stress intensity factor $K_I$ incrementally. For simulations in presence of hydrogen, we fully cover the crack surfaces, as they are the strongest hydrogen traps\textsuperscript{\cite{Carter_2004_PhysRevB.70.064102}}. The crack tip, due to the significant stress triaxiality, is another natural trap\textsuperscript{\cite{Vehoff1997}}.
Therefore, we compute the probability of finding a hydrogen atom at each tetrahedral lattice site around the crack tip for a range of bulk lattice concentrations (0.05 to 200 appm), and allocate hydrogen accordingly. In particular, Oriani statistics is used to determine the distribution of hydrogen atoms at the crack tip. Per concentration, at least three distinct realizations are considered to account for stochastic effects (see Supp. Info. S5 ``Determination of hydrogen distribution at atomistic crack tips'').

We first simulate mode-I loading in pure iron. In the absence of hydrogen, dislocation emission is observed at $K_{Ie}$ = 1.16~MPa$\sqrt{\textup{m}}$. The mechanism initiates by nucleation of a surface step, followed by expansion of an emitted dislocation and propagation of its screw components until their periodic counterparts meet and annihilate. Under the action of the Peach-Koehler force induced by the $K$-field, the straight dislocation eventually glides away from the crack front, leaving behind a blunted crack (see Supp. Video 1).

Next, we consider 5 appm lattice concentration of hydrogen. As the load is increased, more hydrogen is attracted to the crack tip. Additional hydrogen atoms are inserted as a function of the increased $K$ according to Oriani statistics (see Supp. Info. S5 ``Determination of hydrogen distribution at atomistic crack tips''). Eventually, the crack propagates at $K_{Ic}$ = 0.86~MPa$\sqrt{\textup{m}}$ (see Supp. Video 2). Thus, a few appm hydrogen induce a transition from emission to cleavage, i.e. from intrinsically ductile to brittle behavior.

In order to assess the dependence of crack advance on the hydrogen concentration, we next drastically reduce the hydrogen lattice concentration to 0.05 appm.
Although cleavage remains favorable, crack growth requires a higher load ($K_{Ic}$ = 0.96~MPa$\sqrt{\textup{m}}$). This shift suggests that the competition between dislocation emission and cleavage is still in favor of the latter, but not as markedly as at 5 appm. Finally, an increase of the hydrogen content to 200 appm results in cleavage at a very low stress intensity factor of $K_{Ic}$ = 0.76~MPa$\sqrt{\textup{m}}$.

Our results are consistent with experiments. For instance, Vehoff {\it et al.} investigated dislocation activity near crack tips in Fe-2.6\%Si single crystals, with and without hydrogen\textsuperscript{\cite{VEHOFF1980,VEHOFF1986_2nd}}, see Figs.~\ref{proposed_theory}a,b. In the absence of hydrogen, the material is ductile, characterized by blunting and considerable crack-tip plasticity. Upon hydrogen exposure, crack-tip plasticity events plummet, in favor of brittle cleavage that becomes more prevalent as the hydrogen content rises. Based on these experiments, Vehoff {\it et al.} concluded that embrittlement takes place at the crack tip\textsuperscript{\cite{VEHOFF1980}}, as evidenced by the tip's sharpness: a sharper tip indicates a greater share of cleavage compared to emission\textsuperscript{\cite{McClintock1968}}. Recently, Birenis {\it et al.} employed modern experimental techniques and reached similar conclusions\textsuperscript{\cite{BIRENIS2018245,BIRENIS2019396}}. 

\subsection*{\normalsize{\textit{Theory of embrittlement}}}
To put our simulation results into context, we turn to Griffith's theory of brittle fracture. According to this theory, cleavage occurs when the strain energy released by crack growth exceeds the energy required to create two new surfaces, thus favoring cleavage in materials with lower surface energy. Taking into consideration that the surface energy $\gamma_s$ depends on the surface coverage of hydrogen, Griffith's criterion for brittle fracture in the presence of hydrogen takes the form 
\begin{equation}\label{Eq:theory}
    K_{\text{G}} = \sqrt{\frac{2\gamma_{\text{s}}(\Theta)}{B}}\ ,
\end{equation}%
where $B$ depends on the elastic constants (see Supp. Info. S4.2 ``Critical stress intensity factor calculation details'') and $\Theta$ is the ratio between the number of hydrogen atoms and the number of binding sites available on the nucleating fresh surface. Since $K_G$ is a lower bound for atomistic simulations because of lattice trapping, the trapping barrier $\Delta K_\text{trap}$ should be added to Eq.~(\ref{Eq:theory}) to match atomistic simulation results. Here, $\Delta K_\text{trap}=0.15\ \rm MPa\ \sqrt{m}$ is computed for pure Fe as the difference between $K_{Ic}$ and $K_G$ (see Supp. Info. S4.2 ``Critical stress intensity factor calculation details'').

While a reduced surface energy favors cleavage, there needs to be a fast hydrogen supply for the crack to advance. By contrast, oxygen does lower the surface energy, but does not induce embrittlement: this might be due to its slow diffusion in bcc iron\textsuperscript{\cite{hancock1966hydrogenoxygen,Vehoff1997}}. We therefore track the number of iron-iron bonds at the crack tip and the corresponding number of hydrogen atoms on the adjacent crack surfaces during the course of the simulations. Strikingly, the two display near-perfect mirror symmetry when viewed as a function of the applied $K$ (Fig.~\ref{bonds_and_h_atoms}). We interpret this trend as follows: any influx of hydrogen leads to immediate bond breaking; when hydrogen flows out, bonds reform. Note that this observation is confirmed by considering multiple statistical distributions of hydrogen at every concentration (see Supp. Info. S6 ``Hydrogen atoms count at the crack tip and its relation to surface energy''). This consistency verifies hydrogen-induced local decohesion.

Next, we compute the surface energy $\gamma_s(\Theta)$ as a function of the hydrogen coverage that is obtained by measuring the hydrogen content at the nucleating crack surfaces after cleavage (Fig. \ref{bonds_and_h_atoms}), see Supp. Info. S6 ``Hydrogen atoms count at the crack tip and its relation to surface energy'' and Supp. Figs. S4-S6. By using Eq.~(\ref{Eq:theory}) and the trapping barrier $\Delta K_{\text{trap}}$, we compute the critical stress intensity factor $K_{Ic}$. The theory predictions closely match the simulation results (Fig.~\ref{proposed_theory}), thus supporting a Griffith mechanism enabled by rapid hydrogen supply to the cleavage region.

The theory also implies a saturation level: once enough hydrogen accumulates near the crack tip to fully cover the nucleating surfaces, the stress intensity required for cleavage cannot decrease any longer. Indeed, at 200 appm, where surface coverage approaches full
saturation $\Theta\approx1$, the critical stress intensity $K_{Ic}$ levels off. This outcome rationalizes experimental observations of saturation\textsuperscript{\cite{VEHOFF1986_2nd}}.


\section*{\normalsize{Discussion}}

Our findings draw a simple physical picture: embrittlement occurs when fast-diffusing hydrogen reduces surface energy, enabling cleavage at the crack tip. Our simulation results clarify how this process unfolds at the atomic scale, and can be used to inform microscale models of embrittlement. This picture is consistent with currently known experimental observations\textsuperscript{\cite{VEHOFF1980,VEHOFF1986_2nd,BIRENIS2018245,BIRENIS2019396}}, and is remarkably simple in comparison with recent propositions (such as refs. \cite{gabor_theory_SHISHVAN2020103740,Song_Curtin_NatMat_2013}), which require the formation of hydrides at crack tips in order to predict embrittlement.

Here, our direct atomistic simulations using a near-DFT-accurate interatomic potential demonstrate that brittle crack propagation can occur well below the critical stress intensity factor for emission. This result is in contrast with the hypothesis of cleavage induced by hydride formation, which occurs at a critical stress intensity factor that is larger than the one for emission\textsuperscript{\cite{Song_Curtin_NatMat_2013}}.  The modified Griffith criterion that stems from our analysis is instead consistent with previous assessment based on the assumption of thermodynamic equilibrium of the decohering surface with the hydrogen-containing lattice\textsuperscript{\cite{jiang_carter_ActaMat_2004_compute_surf_ener_with_H,Oriani1972}}. In particular, our work demonstrates that hydrogen-assisted crack propagation in iron is essentially an {\it athermal} process due to the fast hydrogen diffusion that promotes debonding: it is the combination of fast diffusion and lowered surface energy that favors crack propagation. This mechanism, as incorporated in Eq. \ref{Eq:theory}, should inform higher (e.g. micro-scale) models including interaction of the propagating crack with other defects\textsuperscript{\cite{Gong2020}} (dislocations, grain boundaries) to assess the impact of hydrogen-assisted cracking on embrittlement.








\printbibliography[title={\normalsize{References}}]

\section*{\normalsize{Methods}}

\subsection*{\normalsize{Density Functional Theory (DFT) calculations}}

For the DFT\textsuperscript{\cite{PhysRev.136.B864-DFT-1,PhysRev.140.A1133DFT-2}} calculations of the elastic constants, energy-volume curves, traction-separation curves, and the reference database expansion, we employed the~{\UrlFont{VASP}} package\textsuperscript{\cite{vasp-1,vasp-2,vasp-3}}  with the same settings as those used by Meng et al. to generate the NNP reference database (Section II in Ref.~\cite{NNP_PRM_Ogata_2021}). Specifically, we performed spin-polarized calculations using the projector augmented wave (PAW) method for pseudopotential\textsuperscript{\cite{PAW}}, the GGA-PBE exchange-correlation functional\textsuperscript{\cite{PBE}}, a Monkhorst-Pack scheme for sampling the Brillouin zone\textsuperscript{\cite{Kmesh}}, and the Methfessel-Paxton method for integration\textsuperscript{\cite{Methfessel-Paxton-PhysRevB.40.3616}}; the cutoff energy was set to 360 eV and k-point density to 0.03 $\Ang^{-1}$. More details regarding the DFT database are in Supp. Info. S2 ``ACE FeH potential database and fitting''.

\subsection*{\normalsize{Interatomic potential training and validation}}
We employed the~{\UrlFont{PACEMAKER}} package\textsuperscript{\cite{pacemaker-Lysogorskiy2021}} for training ACE models. The detailed description, the results of the training, and the link to the relevant input files are reported in Supp. Info. S2 ``ACE FeH potential database and fitting''.

For most validation tests, we developed a Python-based workflow based on the Atomic Simulation Environment (ASE) package\textsuperscript{\cite{larsen2017ase}}. We used the FIRE algorithm\textsuperscript{\cite{FIRE_Bitzek_PhysRevLett.97.170201}} with a default $10^{-2}$ eV/$\Ang$ force criterion in all calculations where atomic positions required relaxation, employed the {\UrlFont{PHONOPY}} package\textsuperscript{\cite{phonopy}} for phonon calculations, and leveraged the nudged elastic band (NEB) method\textsuperscript{\cite{NEB_10.1063/1.1329672}} with a spring constant equal to 0.1~eV/$\Ang$, as it is implemented in ASE to compute the hydrogen diffusion paths.

\subsection*{\normalsize{Molecular dynamics simulations}}

We performed molecular dynamics simulations of the loaded crack tips using the~{\UrlFont{LAMMPS}} package\textsuperscript{\cite{LAMMPS}}, executing all simulations within the canonical (NVT) ensemble. We set a time step of 2 fs for pure Fe and 0.4 fs for Fe with H, which corresponds to roughly 1/45 of the inverse highest phonon frequencies for these systems (see Figs. S2e,f in Supp. Info. Supp. Info. S3 ``Validation of the ACE FeH potential against benchmark properties''). 

\section*{\normalsize{Data Availability Statement}}

Link to the data related to this publication is reported in the Supplementary Information file.

\section*{\normalsize{Code Availability Statement}}

Link to the code related to this publication is reported in the Supplementary Information file.

\section*{\normalsize{Acknowledgements}}

This work made use of the Dutch national e-infrastructure with the support of the
SURF cooperative using grant no. EINF-12585. We thank the Center for Information Technology of the University of Groningen for their support and for providing access to the Hábrók high-performance computing cluster. 

\section*{\normalsize{Funding}}

This research was carried out under project number N19009 in the framework of the Partnership Programme of the Materials Innovation Institute M2i (https://www.m2i.nl) and the Netherlands Organisation for Scientific Research (https://www.nwo.nl) and with the support of Tata Steel Nederland Technology B.V., Allseas Engineering B.V. and Nedschroef Helmond B.V.

\section*{\normalsize{Author Contributions}}
\textbf{Aleksei Egorov:} Conceptualization, Data curation, Formal analysis, Investigation, Methodology, Validation, Visualization, Writing – original draft, Writing – review \& editing. 
\textbf{Lei Zhang:} Conceptualization, Data curation, Formal analysis, Investigation, Methodology, Validation, Visualization, Writing – review \& editing.
\textbf{Erik Van der Giessen:} Conceptualization, Funding acquisition, Methodology, Project administration, Resources, Supervision, Writing – review \& editing.
\textbf{Francesco Maresca:} Conceptualization, Funding acquisition, Methodology, Project administration, Resources, Supervision, Writing – review \& editing.

\section*{\normalsize{Competing Interests Declaration}}

Authors declare no competing interests.

\end{refsection}

\newpage

\clearpage
\appendix
\onecolumn











\setlength\cellspacetoplimit{3pt}
\setlength\cellspacebottomlimit{3pt}

\definecolor{natcommtab1}{RGB}{214,214,194}
\definecolor{natcommtab2}{RGB}{236,235,226}
\definecolor{natcommtab3}{RGB}{251,250,246}

\captionsetup{font=small, labelfont=bf, singlelinecheck=false}





\renewcommand{\thefootnote}{\fnsymbol{footnote}} 

\pagestyle{fancy}
\fancyfoot[C]{\textcolor{gray}{Page \thepage\;of \pageref*{LastPage}}}
\fancyhead{} 
\renewcommand{\headrulewidth}{0pt} 


\renewcommand{\v}[1]{\boldsymbol{\mathbf{#1}}} 
\renewcommand{\l}[0]{\left} 
\renewcommand{\r}[0]{\right} 
\let\f=\frac 
\renewcommand{\t}[1]{\text{#1}} 

\renewcommand{\arraystretch}{1.3} 

\renewcommand{\thefigure}{S\arabic{figure}}
\renewcommand{\theequation}{S\arabic{equation}}
\renewcommand{\thetable}{S\arabic{table}}
\renewcommand{\thesection}{S\arabic{section}}

\captionsetup[table]{font={large, stretch=1.2}}     
\captionsetup[figure]{font={large, stretch=1.2}}    


\newgeometry{top=1in,bottom=1in,left=0.7in,right=0.7in}

\onecolumn
\begin{flushleft}
\Large

Supplementary Information for \\
\huge
\textbf{To crack, or not to crack: How hydrogen favors crack propagation in iron at the atomic scale}

\vspace{0.25cm}


\onehalfspacing

\large
Aleksei Egorov\textsuperscript{1},
  Lei Zhang\textsuperscript{2}, Erik van der Giessen\textsuperscript{1}, and
  Francesco~Maresca\textsuperscript{2}\\
  \vspace{0.05cm}
  \textit{\small \textsuperscript{1}Zernike Institute for Advanced Materials, Faculty of Science and Engineering, University of Groningen, Nijenborgh 3, 9747 AG, Groningen, The Netherlands \\ \small \textsuperscript{2}Engineering and Technology Institute Groningen (ENTEG), Faculty of Science and Engineering, University of Groningen, Groningen 9747AG, the Netherlands} \\

\end{flushleft}

\vspace{-0.1cm}

\large
 
\onehalfspacing
\setcounter{figure}{0}
\begin{refsection}
    

\tableofcontents

\section{\Large Literature on atomistic modelling of cracks with hydrogen}

Atomistic modelling of bcc metals is intrinsically challenging in the case of cracks. On the one hand, empirical (EAM) potentials deliver contradicting or incorrect predictions regarding the intrinsic crack-tip behaviour under mode-I loading, see e.g. Refs. \cite{moller_bitzek_2014comparative_eam_drawbacks} and \cite{zhang_NPJ_2023atomistic} for bcc Fe. On the other hand, machine learning potentials are generally prone to artifacts if not trained on and tested for crack-tip relevant configurations, as discussed in Ref. \cite{zhang_NPJ_2023atomistic}. In particular, Zhang et al.\textsuperscript{\cite{zhang_NPJ_2023atomistic}} have shown by means of a Gaussian Approximation Potential that the T=0K intrinsic fracture behaviour of $\{100\}$ and $\{110\}$ cracks in pure bcc Fe loaded under mode-I is cleavage. This same behaviour is reproduced by a range of machine-learning frameworks, including ACE as used here (see Supp. Info. 2), if a suitable database is used for the training that contains traction-separation data and crack-tip-relevant configurations\textsuperscript{\cite{Zhang2024}}. Finally, suitable boundary conditions, consistent with Linear Elastic Fracture Mechanics (LEFM), should be applied at the boundaries of the molecular dynamics simulation cell\textsuperscript{\cite{andric_curtin_2018atomistic}}, to ensure simulation cell size independence of the results.

Thus, assessment of existing literature should be based on verifying (i) the suitability of the potential for modelling cracks in pure Fe and in presence of H; and (ii) the appropriateness of the chosen boundary conditions. Previous work based on empirical potentials\textsuperscript{\cite{gabor_theory_SHISHVAN2020103740,Song_Curtin_NatMat_2013}} adopt either Ramasubramaniam et al.\textsuperscript{\cite{Rama2009}}, or Wen\textsuperscript{\cite{Wen2021}} potentials, or other formulations where the Fe--H and H--H interactions are adjusted. In all these cases, the Fe part of these potentials is either the Ackland\textsuperscript{\cite{Ackland2004}} or the Mendelev\textsuperscript{\cite{Mendelev2003}} potential. As discussed in Ref. \cite{Moller2018PF}, both Mendelev and Ackland potentials predict, under mode-I conditions, the formation of an artificial planar fault that has local fcc symmetry. Since crack propagation depends on the interplay between fast hydrogen diffusion and decrease of the surface energy, this local structural change is expected to affect the outcome of atomistic simulations of crack tips loaded in the presence of hydrogen. In fact, where these potentials are used, a local phase transition leading to hydride formation is typically observed, which can therefore be considered to be an artifact of the potential. This artifact affects the key conclusions of Refs. \cite{Song_Curtin_NatMat_2013,Xing2017,gabor_theory_SHISHVAN2020103740,Wang2024}. Moreover, in the case of Refs. \cite{Xing2017} and \cite{Wang2024}, the imposed boundary conditions are not consistent with LEFM and hence lead to size-dependent results, as discussed in Ref. \cite{andric_curtin_2018atomistic}. A choice of boundary conditions that leads to size-dependency is also made in Ref. \cite{Asta2022}. Despite using a machine-learning potential, no detailed validation of benchmark properties, such as Supp. Info. 3, is provided. Finally, Meng et al.\textsuperscript{\cite{NNP_PRM_Ogata_2021}} developed a FeH NNP potential, which has intrinsic shortcomings (see main text). Moreover, this potential has not been used for K-test simulations using LEFM boundary conditions as discussed here.

\section{\Large ACE FeH potential database and fitting}
\newrefcontext[labelprefix=S]
\onehalfspacing

Meng {\it et al.} created a DFT reference database for iron with hydrogen, explicitly focusing on embrittlement\textsuperscript{\cite{NNP_PRM_Ogata_2021}}. This database includes 21,928 configurations, including free surfaces, $\gamma$ surfaces, dislocations and deformed bulk (see Supplementary Information of Ref.~\cite{NNP_PRM_Ogata_2021}). Out of this dataset, at the descriptor construction step, PACEMAKER selected 21,917 structures as the starting point for training our atomic cluster expansion (ACE), and we further included some extensions.

\begin{figure*}[tb!!]
  \centering
  \includegraphics[width=0.99999\textwidth]{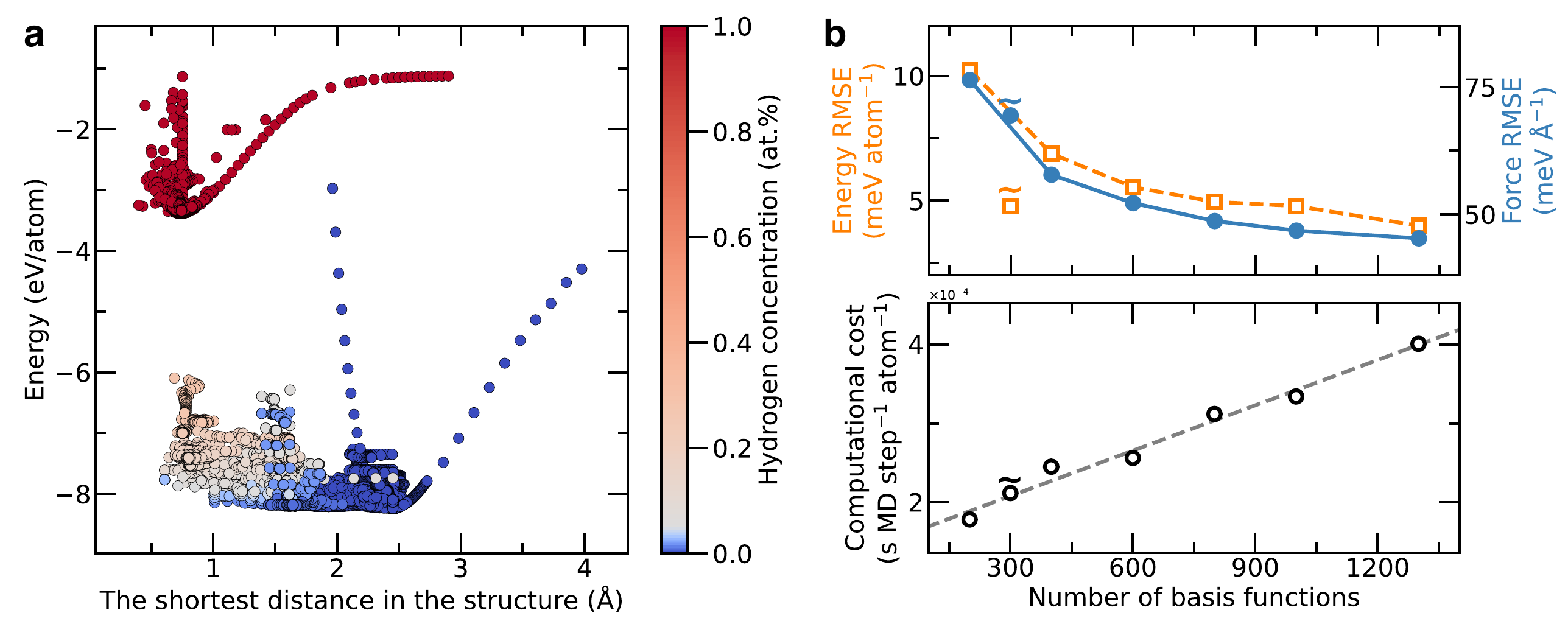}
\caption{\label{dft_db_and_error_vs_complexity} {\textbf{\large Reference data and model complexity vs. accuracy and speed.}
  \large
  \textbf{a}, DFT reference database for ACE training. The database contains 22,305 structures, including 21,917 structures from the NNP database (see ref.~\cite{NNP_PRM_Ogata_2021}), 20 structures from the bcc E-V curve for the low and high volumes (or nearest-neighbor distances), and 368 structures from traction-separation curves. 
  \textbf{b} Accuracy and computational cost versus the complexity of our ACE. By increasing the number of basis functions (bf), accuracy improves in terms of the radial mean square errors (RMSE) for energies and forces for the testing dataset. However, the increased accuracy is accompanied by a higher computational cost, as illustrated by the incremental rise in the number of bf, reaching 1,300 in one of the trained models. Based on the trade-off between accuracy and computational cost, we choose an ACE with 300 b.f. for our final model (denoted by labels with curly hats). 
  }}
\end{figure*}

Additional data include 388 configurations. These additional configurations comprise 20 structures along bcc energy versus volume (E-V) curve, including the small-volume part to ensure a strong repulsion (cf.~Fig.~2b in the main text) and the large-volume part to prevent spurious local minima. Given our focus on cracks, we added 368 structures from the traction-separation (T-S) curves for $\{100\}$ and $\{110\}$ surfaces, including randomly distorted T-S curves to diversify the atomic environments. To produce the T-S curves, a bulk crystal is sliced along a desired plane, and two pieces are rigidly pulled apart. Since DFT cannot handle cracks, T-S curves are the closest proxy\textsuperscript{\cite{curtin_ts_curve_explained_2005methods,Möller2018_ts_explained}}.
Figure~\ref{dft_db_and_error_vs_complexity}a depicts the final database. The database was split into training and testing datasets, with 10\% of the structures randomly selected for testing and the remaining 90\% used for training.

ACE formalism enables us to reduce training/testing errors by increasing the model complexity, reflected by the number of basis functions (bf)\textsuperscript{\cite{Drautz-2019-PhysRevB.99.014104},\cite{Anton-2022-PhysRevMaterials.6.013804}}. However, the computational cost increases linearly with the number of basis functions (Fig.~\ref{dft_db_and_error_vs_complexity}b). Given this trade-off, we opted for 300 basis functions for the final model---while displaying errors on par with NNP, it is $\sim 15$ times faster (see Table~\ref{tab:performance} and Supp. Info. 2.1). 

\begin{table*}[htb]
    \centering
    \begin{threeparttable}
        \large  
        \caption{\textbf{FeH atomic cluster expansion (ACE) errors compared to neural network potential (NNP)\textsuperscript{\cite{NNP_PRM_Ogata_2021}}}. We report root mean square (RMS) errors in meV atom$^{-1}$ and eV $\mathring{\text{A}}$$^{-1}$ for the test data set. *~--~NNP's RMS errors are for the original reference database from Ref.~\cite{NNP_PRM_Ogata_2021}, while ACE's errors are for a larger, extended database (Fig.~\ref{dft_db_and_error_vs_complexity}a); values in the brackets are the errors for the structures within 1~eV from the ground state.}
        \label{tab:performance}
        \renewcommand{\arraystretch}{2.0}  
        \tabcolsep=0.0cm
        \begin{tabular}{@{}l l l@{}}
            \hline
            \rowcolor{natcommtab1} \textbf{\hspace{0.2cm}Potential\hspace{1.8cm}} & \textbf{\hspace{0.2cm}RMS error, energies*, train / test\hspace{1.05cm}} & \textbf{\hspace{0.2cm}RMS error, forces*, train / test\hspace{1.05cm}} \\
            \hline
            \rowcolor{natcommtab3} \hspace{0.2cm}ACE (300 b.f.)\hspace{0.2cm} & 
            \hspace{0.2cm}3.20 (3.04) / 4.78 (2.90)\hspace{0.2cm} & 
            \hspace{0.2cm}69.4 (67.9) / 69.4 (68.3)\hspace{0.2cm} \\
            \hline
            \rowcolor{natcommtab2} \hspace{0.2cm}NNP (ref.~\cite{NNP_PRM_Ogata_2021})\hspace{0.2cm} & 
            \hspace{0.2cm}2.98 / 3.01\hspace{0.2cm} & 
            \hspace{0.2cm}69.7 / 68.2\hspace{0.2cm} \\
            \hline
        \end{tabular}
    \end{threeparttable}
\end{table*}

We developed the final 300-bf ACE in two stages. First, we trained a 200-bf model using a relative energy weight of 0.97 (with respect to forces). This model then served as the starting point for the 300-bf ACE, where we increased the energy weight to 0.998 and extended the number of BFGS optimization steps from 800 to 3500. This setup, refined through extensive trial and error, produced the lowest overall numerical errors, significantly improving energy accuracy with only a minor trade-off in force error (Fig.~\ref{dft_db_and_error_vs_complexity}b).

All other ACE models in Fig.~\ref{dft_db_and_error_vs_complexity}b utilized the same 200-bf potential as a starting point. For these models, we adjusted the energy weights as follows: 0.992 (400 bf), 0.994 (600 bf), 0.995 (800 bf), 0.996 (1000 bf), and 0.998 (1300 bf). Each was trained with 1000 BFGS steps. We fit these models hierarchically, each time using the previous model as input to train the next one: 200~$\rightarrow$~400~$\rightarrow$~600~$\rightarrow$~800~$\rightarrow$~1000~$\rightarrow$~1300~bf.

All models were constructed with a cutoff radius of 6.5 $\Ang$ for Fe-Fe interactions and 3.5 $\Ang$ for Fe--H and H--H, reflecting the shorter range of hydrogen-related bonding. We applied a maximum body order of five (i.e. six-body interactions). The number of radial functions and spherical harmonics body order is set to (26, 8, 3, 2, 1) and (0, 7, 3, 2, 1), respectively---this choice balanced model accuracy and computational cost.

All potentials were non-linear and utilized the Finnis-Sinclair embedding function with standard parameters recommended in the PACEMAKER documentation.

The complete input file for the 300-bf ACE model is included in the Supplementary files. Input files for all other models use the same parameters unless otherwise stated.

\subsection{Computational speed calculation details}

To evaluate the computational cost of the ACE and NNP, we conducted 100 MD simulation steps in the NVT ensemble at $T=100~K$ across three system sizes: $8\times8\times7$, $16\times16\times14$, and $32\times32\times28$, based on the bcc conventional unit cell, with $896$, $7,168$, and $57,344$ iron atoms, and randomly placed 2, 16, and 128 hydrogen atoms, respectively (Fig. 2a of the main text).
Given that at $T=100~K$ the thermal expansion is minimal\textsuperscript{\cite{therm_exp_dft_HAFEZHAGHIGHAT2014274}}, we used $T=0~K$ lattice constants.

We assessed the ACE computational cost for different numbers of basis functions (Fig.~\ref{dft_db_and_error_vs_complexity}b) using AMD 7763 CPU for the smallest system of 896+2 atoms. Because of the higher memory requirements of NNP, which exceeded the capacity of the AMD 7763 CPU, the rest of the simulations (Fig.~2a of the main text) were performed on AMD Rome 7H12 (2.6GHz, 280W).

\vspace{-0.2cm}

\section{\Large Validation of the ACE FeH potential against benchmark properties}\label{Validation_ACE}

\subsection{Validation results}

In this Section, we present validation tests of our ACE for iron with hydrogen and pure iron, in addition to tests illustrated in Fig. 2 in the main text. For all tests, we compare the ACE results with those from DFT and NNP\textsuperscript{\cite{NNP_PRM_Ogata_2021}}. We need to note that the authors of the NNP developed a novel potential (also neural network) that, while exhibiting slightly higher errors than the first version, enables acceleration on GPU, resulting in a speed-up of 40 for molecular dynamics (MD) simulations\textsuperscript{\cite{NNP_2_ZHANG2024112843}}. We compare our ACE with the first NNP\textsuperscript{\cite{NNP_PRM_Ogata_2021}} because it has lower numerical errors. GPU speed-up is irrelevant since ACE also supports MD simulations on GPUs, and Qamar {\it et al.} demonstrated that it can achieve a 60-fold speed-up\textsuperscript{\cite{Qamar2023}}.

\begin{figure*}[!ht]
  \centering
  \includegraphics[width=0.9999999\textwidth]{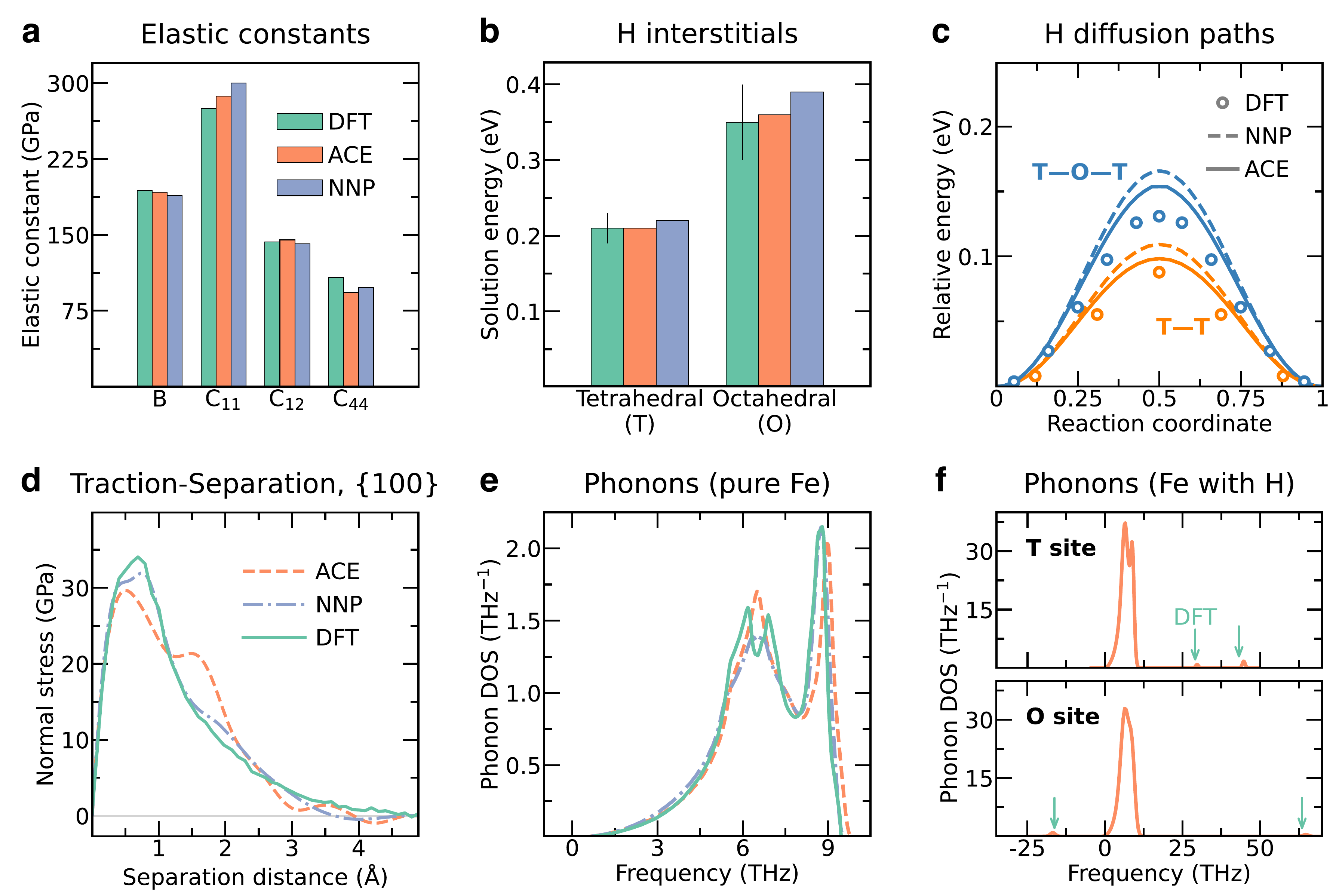}
  \caption{\label{valid_tests} {\textbf{\large Fe-H Atomic Cluster Expansion (ACE) validation tests for bcc Fe.}
  \large
  \textbf{a}, Elastic constants.  
  \textbf{b}, Solution energies of the hydrogen interstitial atoms for tetrahedral (T) and octahedral (O) sites in the bcc lattice.
  \textbf{c}, Energy profiles of diffusion paths between different interstitial sites. DFT data from ref.~\cite{diffusion_paths_Hirata2018}.
  \textbf{d},  Traction-Separation curve for the 100 surface.
  \textbf{e}, Phonon density of states (DOS) for pure iron (DFT data from ref.~\cite{zk_liu_dft_phonons_bcc_fe_HANG201494}). 
  \textbf{f}, Phonon density of states (DOS) for iron with interstitial hydrogen at the tetrahedral (T) and octahedral (O) sites obtained with ACE; the positions of the peaks due to hydrogen are compared to the DFT results from ref.~\cite{phonons_with_h_dft_ALVAREZ2024107590}. DFT solution energies are presented as the average values from refs.~\cite{Carter_2004_PhysRevB.70.064102}, \cite{jiang_carter_ActaMat_2004_compute_surf_ener_with_H}, \cite{wolverton_h_sol_energies_COUNTS20104730}, and \cite{sol_ener_ZHU202238445} with 95\% confidence intervals.
  }}
\end{figure*}

For pure iron, we computed (1) elastic constants, as they impact stress fields around cracks; (2) lattice vibrations---phonons---to ensure our potential behaves plausibly at finite temperatures; and (3) the fracture-related traction-separation (T-S) curve for the $\text{\{}100\text{\}}$ surface (in addition to the $\text{\{}110\text{\}}$ surface presented in the main text). Both ACE and NNP align with DFT for these three tests (Fig.~\ref{valid_tests}a,d,e).

In order to interact with the crack tip at an applied load, hydrogen atoms must first diffuse towards it and bind to the relevant (tetrahedral) lattice sites. We therefore compute the solution energies of the hydrogen atoms at two interstitial sites---tetrahedral (T) and octahedral (O) sites---and evaluated the two transition paths between them. Both the former and the latter align with DFT (Fig.~\ref{valid_tests}b,c). Therefore, in finite-temperature simulations hydrogen is expected to diffuse through the iron lattice at a realistic rate and to follow an accurate path.

Furthermore, in order to determine the appropriate MD timestep for simulations of iron with hydrogen, which depends on the highest frequency vibrational modes\textsuperscript{\cite{MD_timestep_KIM201460}}, we calculated the phonon density of states (DOS) for bcc iron with interstitial hydrogen at both the T and O sites (Fig.~\ref{valid_tests}f). The resulting timestep is 0.4 fs, which corresponds to approximately 1/45 of the inverse of the highest phonon frequency. These results also serve as an additional validation test, as ACE-DOS aligns closely with the DFT results from ref.~\cite{phonons_with_h_dft_ALVAREZ2024107590}.

Finally, to conduct crack tip simulations at 500 K, we adjusted the lattice parameter based on the thermal expansion coefficient at this temperature. The value of 1.02 obtained with ACE is consistent with the DFT result of 1.015\textsuperscript{\cite{therm_exp_dft_HAFEZHAGHIGHAT2014274}}.

\vspace{-0.2cm}

\subsection{Validation setups}

The elastic constants were computed using the \texttt{amstools} package\textsuperscript{\cite{amstools}}. The bulk modulus was obtained by fitting the Murnaghan equation of state to 25 data points from the energy-volume curve within the volume range of 0.9 to 1.1, with 1.0 corresponding to the equilibrium volume.

We employed a 3$\times$3$\times$3 conventional unit cell containing 54 iron atoms to determine the solution energies for interstitial hydrogen atoms. Subsequently, we utilized the same relaxed cell for the nudged elastic band (NEB) calculations\textsuperscript{\cite{NEB_10.1063/1.1329672}} of the diffusion pathways.

The traction-separation (T-S) curves were generated by slicing the bulk cell along the relevant plane and then rigidly separating the two parts. Slabs composed of eight layers of iron atoms were utilized, ensuring the T-S curves are converged. The slight oscillations in the DFT T-S curve come from the \texttt{VASP} settings; increasing the cutoff energy can eliminate them. However, we maintained Meng {\it et al.} original \texttt{VASP} settings\textsuperscript{\cite{NNP_PRM_Ogata_2021}} to ensure consistency with their reference database.

Phonons were calculated using the workflow in the \texttt{amstools} package\textsuperscript{\cite{amstools}}. We employed a 3$\times$3$\times$3 conventional unit cell containing 54 iron atoms to determine the phonon density of states (DOS) of bcc iron with interstitial hydrogen atoms. Before calculating the phonons, the atomic positions and cell volume were relaxed.

Our workflow for the validation tests is provided in the accompanying Jupyter Notebook.

To determine the thermal expansion coefficient and the lattice parameter at $T=500~K$, we conducted an MD simulation employing a 5$\times$5$\times$5 conventional unit cell containing 250 iron atoms in the NPT ensemble for 60 ps. We calculated the average lattice parameter after discarding the initial 6 ps to allow the system to equilibrate\textsuperscript{\cite{freitas-therm-exp}}.

\vspace{-0.2cm}

\section{\Large Crack tip simulation setup}

\subsection{Choice of the converged simulation cell size}

\begin{figure*}[!ht]
  \centering
  \includegraphics[width=0.99999\textwidth]{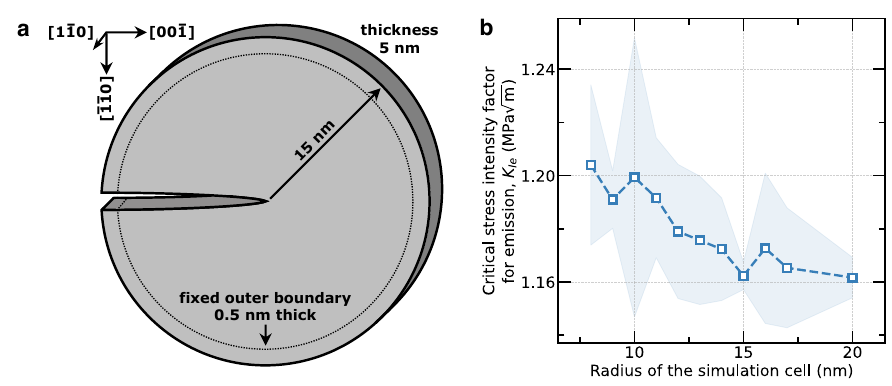}
  \caption{\label{sim_cell_and_conv} {
  %
  \large
  \textbf{a}, Schematic of the simulation cell for the $(\bar{1}\bar{1}0)[1\bar{1}0]$ crack system, employed in this work. The cell contains 324,766 iron atoms. Iron atoms at the 0.5 nm thick outer boundary are fixed to impose the desired value of the stress intensity factor, $K$.
  \textbf{b}, The critical stress intensity factor for dislocation emission, $K_{Ie}$, as a function of the simulation cell radius. The results are presented as mean values of three simulations with 95\% confidence intervals. The value of $K_{Ie}$ is converged at a radius of 15 nm.
  }}
\end{figure*}

Crack tip simulations are performed using a cylindrical simulation cell and boundary conditions for a semi-infinite crack (Fig.~\ref{sim_cell_and_conv}a). Andric and Curtin demonstrated that this geometry, with a sufficiently large cell radius, is consistent with linear elastic fracture mechanics (LEFM)\textsuperscript{\cite{andric_curtin_2018atomistic}}. To prevent the crack from healing at finite temperatures, we employed a so-called \textit{screening}, by switching off the interactions between iron atoms on opposite crack surfaces\textsuperscript{\cite{andric_curtin_2018atomistic}}.

The simulation cell radius must be sufficiently large to minimize interaction between the crack tip, or the dislocation it emits, and the boundary. We therefore tested the radius convergence of the stress intensity factor for dislocation emission, $K_{Ie}$, where $K_{Ie}$ is a material property and thus should remain constant for any converged radius. The results of the convergence test indicate that a radius of 15~nm is converged (Fig.~\ref{sim_cell_and_conv}b), hence we use this radius for our simulations. Additionally, the cell must be thick enough for the dislocation half-loop to fully form in the event of blunting and for hydrogen atoms to diffuse along the crack tip. For these reasons, we used a simulation cell of approximately 5.2 nm, which corresponds to 13 periodic units in the [1$\bar{1}$0] direction.

We applied a stress intensity factor to the crack tip by adjusting the atomic positions according to a protocol developed by Andric\textsuperscript{\cite{andric_phd_diss_2019mechanics}}. During the simulations, we increased the stress intensity factor by 0.001 MPa$\sqrt{\textup{m}}$ every 2 ps, resulting in a loading rate of 5$\times$$10^{8}$ MPa$\sqrt {\textup{m}}$/s. The influence of the loading rate on the simulation results is discussed in Section \ref{Sec:loading_rate}.

The typical simulation time was approximately 0.25~ns. For example, in the case of 200 appm (rightmost panels in Fig.~\ref{fig:all_graphs}), we initiated simulations at $K$~=~0.65~MPa$\sqrt{\textup{m}}$. Cleavage occurred after about 0.20-0.22~ns, ensuring sufficient time for atomic relaxation and equilibration. After cleavage, we continued the simulation for an additional 0.03-0.05~ns to allow hydrogen saturation on the newly formed crack surfaces.

It is worth noting that Meng {\it et al.}\textsuperscript{\cite{NNP_PRM_Ogata_2021}} recently utilized their neural network potential to simulate crack tips in iron with hydrogen, observing a transition from emission to cleavage\textsuperscript{\cite{NNP_2_ZHANG2024112843}}. However, they employed a nanoscale center-crack tension (CCT) simulation cell geometry. As Andric and Curtin demonstrated, a nano-sized CCT geometry does not adhere to linear elastic fracture mechanics (LEFM), compromising the reliability of simulation predictions\textsuperscript{\cite{andric_curtin_2018atomistic}}. Additionally, Meng {\it et al.} did not provide a cell size convergence test and used a thin cell of 0.8 nm. Lastly, they did not explore the underlying causes of embrittlement.




\subsection{Critical stress intensity factor calculation details}

Eq. 1 in the main text uses the factor $B$ as an input parameter. $B$ is derived from elastic constants that account for anisotropy (see details in ref.~\cite{zhang_NPJ_2023atomistic}). Another input for Eq. 1 is lattice trapping, $\Delta K_{\text{trap}}=0.15\ \rm MPa\ \sqrt{m}$. We determined it as the difference between the Griffith critical stress intensity factor for pure bcc iron (first term in Eq.~1) and the critical stress intensity factor obtained in MD simulations using the Fe ACE potential.

\section{\Large Determination of hydrogen distribution at atomistic crack tips}

To obtain a realistic distribution of hydrogen at the crack tip, we apply Oriani's theory, which utilizes Fermi-Dirac statistics to relate the bulk hydrogen concentration to the concentration at the tip\textsuperscript{\cite{oriani1970diffusion}}. According to this theory, the probability of a hydrogen atom occupying a given tetrahedral (T) site at the crack tip, $\theta^{i}_c$, can be obtained from
\begin{align*}
    \frac{\theta^{i}_c}{1-\theta^{i}_c} = \frac{\theta^{i}_l}{1-\theta^{i}_l} \exp{\left( -\frac{\Delta E^i_b}{k_b T}\right)}\ ,
\end{align*}
where $\theta^{i}_l$ is the probability of a hydrogen atom occupying the T site in a perfect bcc lattice; given the significantly higher stability of T sites compared to octahedral (O) [Fig.~\ref{valid_tests}b], $\theta^{i}_l$ can be regarded as the bulk hydrogen concentration. $k_b$ and $T$ are the Boltzmann constant and the temperature, respectively. $\Delta E^i_b$ denotes the binding energy of a hydrogen atom at the T site.

We calculated the binding energy of hydrogen to the crack tip using elasticity theory (analogous to the binding energy of dislocations and solutes\textsuperscript{\cite{CLOUET20083450_elast_theor_disl_sol_interact}}), which is expressed as
\begin{align*}
    \Delta E_b = V_0 \sigma^{\rm crack}_{ij} \varepsilon^{\rm misfit}_{ij}\ ,
\end{align*}
where $V_0$ is the volume of an iron atom in the bulk bcc lattice. The stress tensor \(\sigma^{\rm crack}_{ij}\) is derived from anisotropic elasticity (see details in ref. \cite{zhang_NPJ_2023atomistic}). The misfit strain \( \varepsilon^{\text{misfit}}_{ij} \) accounts for the deformation caused by introducing a single hydrogen atom into the lattice and is expressed as
\begin{align*}
    \sigma^{\rm crack}_{ij} = \frac{1}{V_0} S_{ijkl} P_{kl}\ ,
\end{align*}
where \( S_{ijkl} \) is the anisotropic compliance tensor of bcc iron and \( P_{kl} \) is the elastic dipole tensor. 

A hydrogen atom was introduced into a $4\times4\times4$ bcc supercell to compute the elastic dipole tensor, \( P_{kl} \). The atomic positions were then relaxed while maintaining fixed lattice vectors. Relaxation induces internal stresses in the supercell, from which the dipole tensor \( P_{ij} \) can be derived as
\begin{align*}
    P_{ij} = -V \sigma_{ij}\ ,
\end{align*}
where $V$ is the volume of the supercell. The corresponding stress tensor of the $\rm F_{128}H$ cell is diagonal, with components $\sigma_{11}=\sigma_{22}=-0.785\rm\ GPa$ and $\sigma_{33}=-0.733\rm\ GPa$. The corresponding misfit tensor $\Omega_{ij}=V_{0}\varepsilon_{ij}^{\rm misfit}$ is also diagonal, with components $\Omega_{11}=\Omega_{22}=2.107\ \Ang^3$ and $\Omega_{33}=1.572\ \Ang^3$, resulting in a total misfit volume $\Delta \Omega = 5.786\ \Ang^3$.

The description above does not account for hydrogen-hydrogen interactions. According to DFT, hydrogen atoms exhibit strong repulsion when located in the nearest T sites in bcc iron, making such a configuration unstable\textsuperscript{\cite{counts2011binding}}. For this reason, in our simulations, we arranged hydrogen atoms around the crack tip based on the probabilities $\theta^{i}_c$ and subsequently removed one hydrogen atom from any hydrogen pairs occupying adjacent T sites. This results in $10--15\%$ less hydrogen atoms than the target local concentration. We also note that the interaction energies for more distant configurations are small\textsuperscript{\cite{counts2011binding}}, and hence we did not consider them.

As described above, we initiated the simulation by setting a specific hydrogen distribution at the crack tip. As the applied load, represented by the stress intensity factor, increases during the simulation, the tensile stresses at the crack tip also rise\textsuperscript{\cite{RITCHIE2021_chap_3}}, thereby increasing the binding energy, $\Delta E^i_b$. Consequently, the crack tip attracts additional hydrogen atoms. To account for this effect, we calculated the hydrogen concentration around the crack tip at two distinct stress intensities: the initial one, and another one that is 0.1~MPa$\sqrt{\textup{m}}$ greater. We then linearly approximated the average number of hydrogen atoms at the crack tip for each increment of 0.001~MPa$\sqrt{\textup{m}}$ in stress intensity. For the three lattice concentrations considered in this work (0.05, 5, and 200 appm) the hydrogen atoms added to the crack tip were 0, 1 and 5, respectively The implementation of H addition is in the accompanying {\UrlFont{LAMMPS}} input file.

Surfaces are the strongest hydrogen traps, and we allocated most hydrogen atoms to cover the crack surfaces. In order to achieve full surface coverage, we let hydrogen atoms adsorb after gradually inserting hydrogen molecules between the crack surfaces. The adsorption process was simulated by fixing all atoms except for the $H_2$ molecules, including iron atoms and hydrogen at the crack tip, and equilibrated the system in the NVT ensemble for 0.6 fs at $T=500~K$. During this period, most hydrogen atoms adhered to the crack surfaces. The process was repeated until both crack surfaces were fully covered. Finally, we removed the remaining $H_2$ molecules between the surfaces to prevent pressure build-up from the hydrogen gas. The accompanying \texttt{LAMMPS} input files provide the detailed implementation.

\vspace{-0.2cm}

\section{\Large Hydrogen atoms count at crack tip and its relation to surface energy}

\begin{landscape}
\begin{figure*}[p] 
  \centering
  \includegraphics[width=1.365\textwidth]{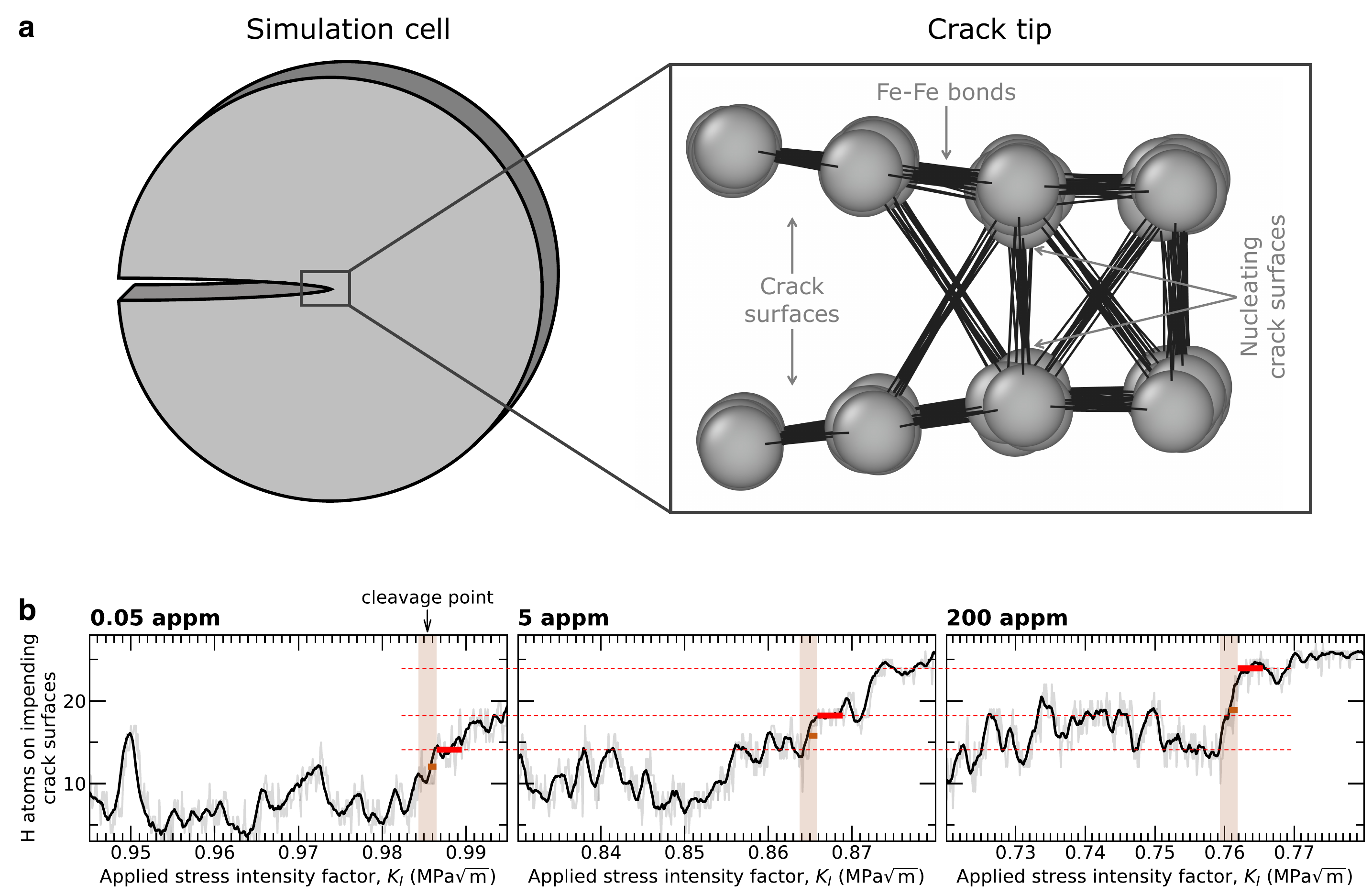}
  \caption{\label{schematic_calc_h_atoms} {
  \large
  \textbf{a}, A schematic illustrating the crack tip geometry, Fe-Fe bonds, and the definition of the nucleation of crack surfaces. \textbf{b}, A few illustrative examples of the evolution of the number of hydrogen atoms on the nucleating crack surfaces as a function of the applied stress intensity factor, $K_{I}$, for various hydrogen concentrations. The shaded vertical bands mark the cleavage point, which is characterized by a sharp increase in hydrogen content.  Thick red strips denote the time-averaged number of hydrogen atoms over an 8 ps window, 2 ps after cleavage (thick brown strips).
  }}
\end{figure*}
\end{landscape}

\begin{figure}[p]
  \centering 
  \begin{overpic}[width=\linewidth]{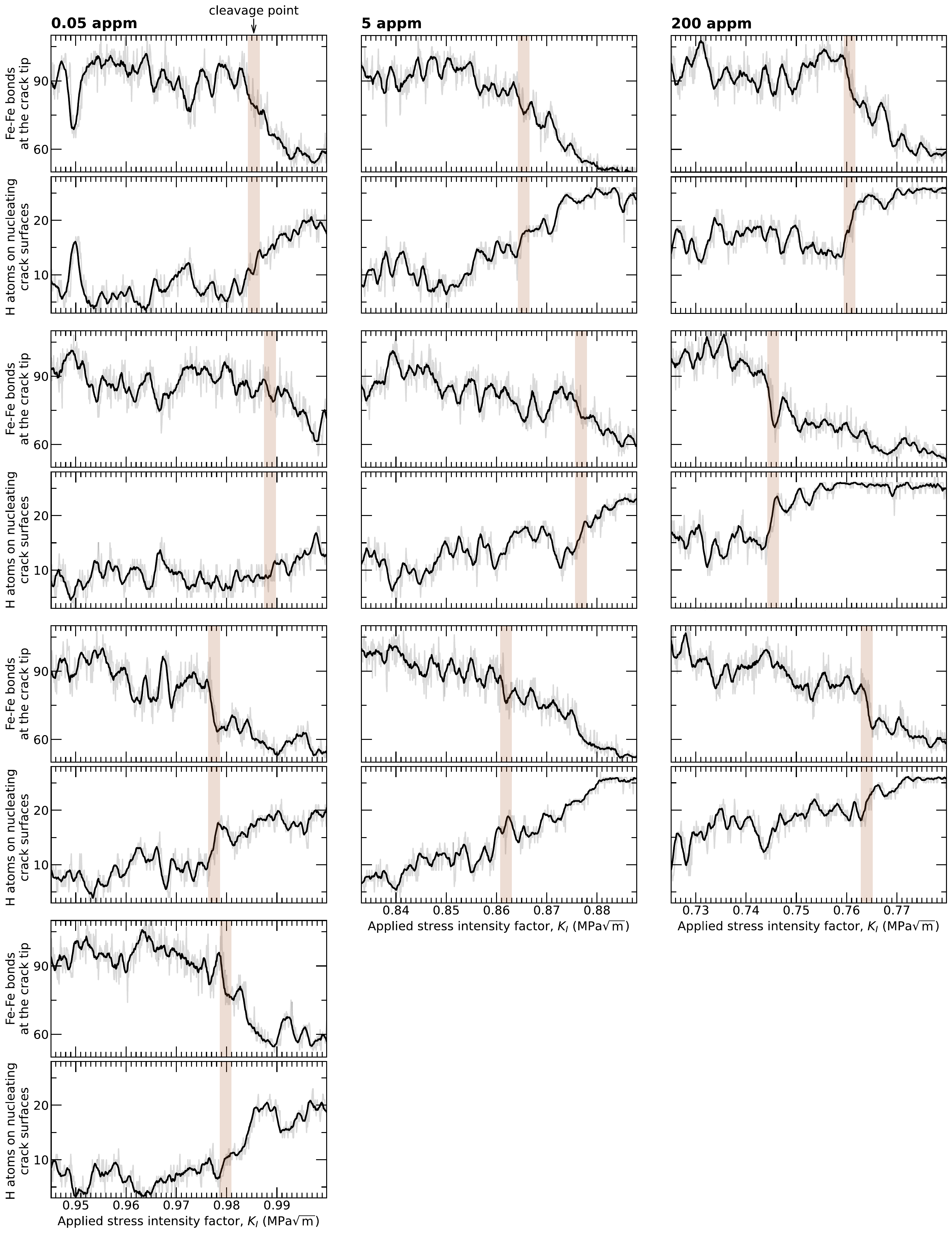}
    \put(29.5,10.88){\parbox{0.605\linewidth}
    {\begin{spacing}{1.5}\large\textbf{Figure S5.}  \textbf{The number of iron-iron chemical bonds at the crack tip and the number of hydrogen atoms on the nucleating crack surfaces (addition to Fig. 3 of the main text).} The results are retrieved from crack tip molecular dynamics simulations with ACE for various hydrogen bulk concentrations, depicted as a simulation progression. Pale gray lines represent the actual data, while the thick black lines indicate the moving averages. Cleavage points are marked where there is a sharp drop in the number of iron-iron bonds and are identified automatically by employing convolution with a step function\textsuperscript{\cite{convolution_smith1997scientist}}.
    \end{spacing}
    }}  
  \end{overpic}
  {\captionsetup{labelformat=empty}\caption[]{\label{fig:all_graphs}}}  
\end{figure}

To obtain the hydrogen content just ahead of the crack tip (Fig. \ref{schematic_calc_h_atoms}a) we examined the rapid local increase in the number of hydrogen atoms {\it right after} cleavage. Figs. \ref{schematic_calc_h_atoms}b and \ref{fig:all_graphs} illustrate that this increase is a hallmark of crack propagation across the whole range of concentrations considered in this study. Notably, while {\it at the point of} cleavage, the number of hydrogen atoms increases when moving from 0.05 to 5 appm lattice concentration, it then saturates and remains almost unchanged at 200 appm compared to 5 appm. This saturation occurs due to the limited number of interstitial (tetrahedral) sites ahead of the crack tip. Once these sites are occupied, further increases in hydrogen concentration must be accommodated at other positions surrounding the crack tip (subsurfaces, deeper sites in front of the crack tip, etc.). However, at the onset of cleavage, fast-diffusing hydrogen diffuses rapidly from these positions to the newly formed surfaces because they are energetically favorable and provide more binding sites than the bulk. Here, we observe a significant difference between 5 and 200 appm: even though the number of hydrogen atoms is nearly the same ahead of the crack surfaces, there are many more hydrogen atoms in the surroundings of the crack tip in the case of 200 appm, resulting in a stronger inflow (cf. thin red dashed lines in Fig. \ref{schematic_calc_h_atoms}b). Therefore, the main factor driving cleavage is the amount of hydrogen resulting from this inflow.

We obtained these levels by computing the average number of hydrogen atoms right after the cleavage point. Namely, we computed the level 2 ps post-cleavage, based on a 8 ps window (schematically indicated by the thick brown strips). While these parameters are somewhat arbitrary, minor variations have a minor impact on the final results depicted in Fig. 1 in the main text.

The next step is to translate the average number of hydrogen atoms into surface coverage and calculate the corresponding surface energy. Fig. 2c of the main text presents the $\{110\}$ surface energies for the clean surface and the surface fully covered with hydrogen (with the ACE values aligned with DFT). For intermediate coverages, DFT predicts that the surface energy changes almost linearly\textsuperscript{\cite{jiang_carter_ActaMat_2004_compute_surf_ener_with_H}}. Therefore, we approximate the energies of fractional coverages linearly. To obtain the coverage, we divide the average number of hydrogen atoms (obtained as described in the previous paragraph) by 26, which is the number of sites available on the two fresh surfaces. The computed energies serve as input for Eq. 1 in the main text.


\newgeometry{top=0.2in,bottom=0.2in,left=0.2in,right=0.5in}  
\onecolumn

\clearpage  
\restoregeometry  




\onecolumn
\newgeometry{top=0.01in,bottom=0.6in,left=0.03in,right=0.07in}  

\thispagestyle{empty}

\onecolumn

\clearpage  
\restoregeometry  


\onecolumn
\subsection{Influence of loading rate on crack propagation}
\label{Sec:loading_rate}

\onehalfspacing

In atomistic crack tip simulations, loading rates are several orders of magnitude higher than in (quasi-static) experiments\textsuperscript{\cite{TEHRANCHI_curtin_review_2019106502}}, which limits the time available for thermally activated processes, such as dislocation emission or hydrogen diffusion. However, by varying the loading rate, we can identify convergence trends and determine the rate that yields a physically plausible picture.

We observed that as the applied stress intensity factor increases, the number of iron-iron bonds at the crack tip decreases, indicating cleavage. At the highest loading rate of 5$\times$$10^{10}$~MPa$\sqrt {\textup{m}}$/s, the cleavage response is gradual and diffuse. A 10 times slower loading rate (5$\times$$10^{9}$~MPa$\sqrt {\textup{m}}$/s) yields earlier cleavage and a more pronounced increase in hydrogen concentration. By decreasing 5 times the loading rate, at $10^{9}$~MPa$\sqrt {\textup{m}}$/s, cleavage becomes more abrupt, and it corresponds to a marked increase in hydrogen content. By further halving down the loading rate, the critical stress intensity factor for cleavage does not change considerably and the same trend is observed: hydrogen saturates the newly nucleated surface. Based on these results, we selected a loading rate of 5$\times$$10^{8}$~MPa$\sqrt {\textup{m}}$/s, which represents a converged regime. In fact, it appears that the loading rate should accommodate fast hydrogen diffusion, which is the driver for the lowered critical stress intensity factor for cleavage.

\setcounter{figure}{5}

\begin{figure*}[!tb!!]
  \centering
  \includegraphics[width=0.99999\textwidth]{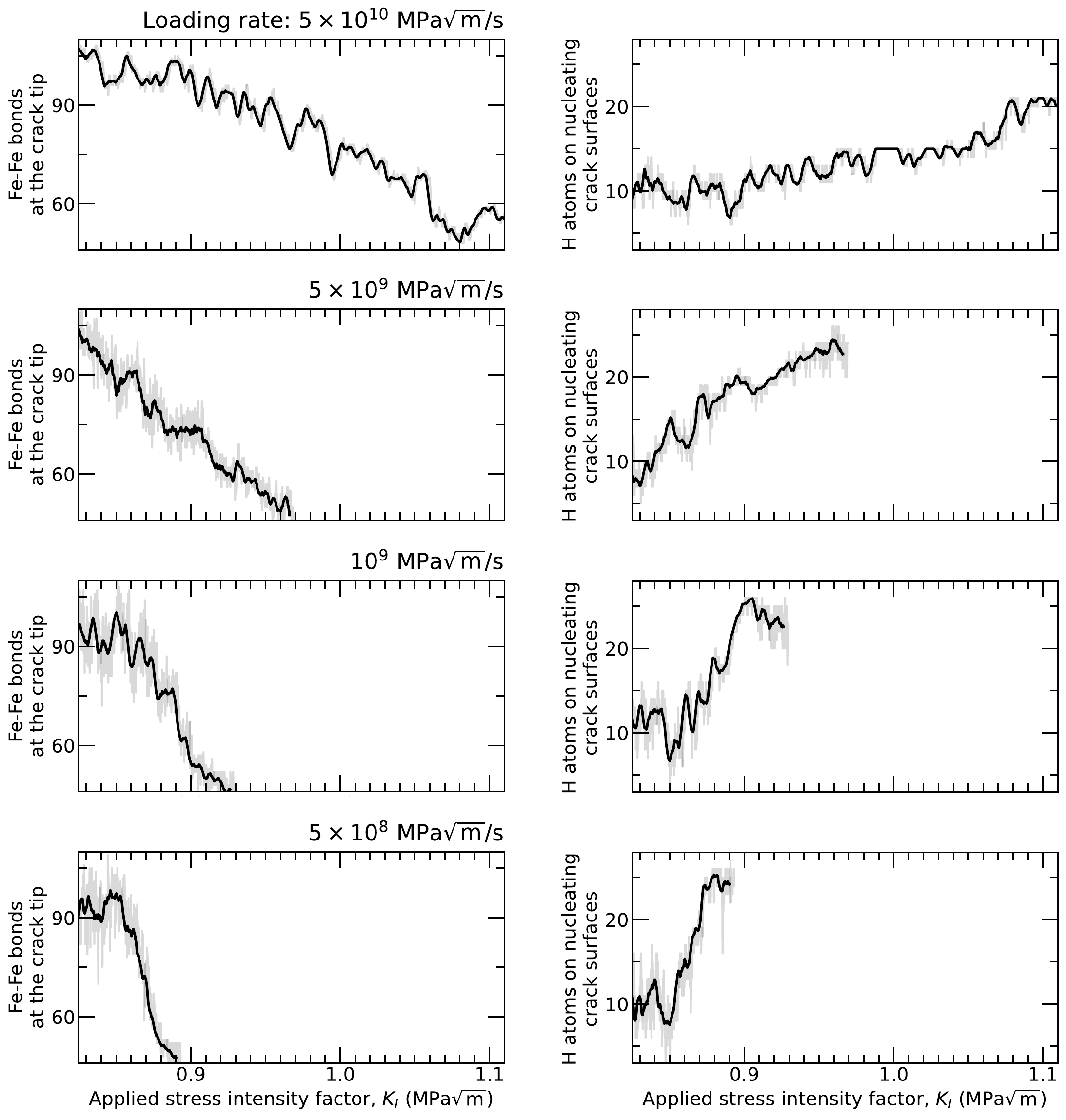}
  \caption{\label{diff_loading_rates} {
  %
  \large
  \textbf{\large Influence of loading rate on crack tip fracture behavior.}
  The number of Fe-Fe bonds at the crack tip versus applied stress intensity factor ($K_I$) at various loading rates, and the corresponding number of hydrogen atoms on nucleating crack surfaces. Lower loading rates lead to more abrupt bond breaking and faster hydrogen accumulation.
  }}
\end{figure*}

To count bonds, we employed the \texttt{OVITO} package\textsuperscript{\cite{OVITOstukowski2009visualization}}, setting a cutoff distance of 3.25~$\Ang$, beyond which cohesion approaches zero (Fig. 2d of the main text). 

\section{\Large Supplementary videos.}

For this pre-print, Supplementary Videos are available upon request from the corresponding author. Supplementary videos 1 and 2 illustrate the simulations described in the main text, created using the OVITO package. Iron atoms are represented in gray, while hydrogen atoms are shown in gold. A cutoff distance of 3.25 $\Ang$ is applied to determine iron-iron bonds, as cohesion becomes negligible beyond this distance (see Fig. 2d in the main text). For pure iron (Video 1), we utilized the dislocation analysis feature of OVITO to visualize the dislocation emitted as a bright green line. In the "Top view" of Video 1, iron atoms are removed for clarity, and the defect mesh, along with the dislocation, is displayed.

\paragraph{\large Supplementary Video S1. Side view.} Dislocation emission process in pure Fe at T=500K. Iron atoms are gray, bonds are the black lines. The emitting dislocation line is highlighted in green.

\paragraph{\large Supplementary Video S1. Top view.} Dislocation emission process in pure Fe at T=500K. Only the crack front and the emitting dislocation line are visualized.

\paragraph{\large Supplementary Video S2.} Crack propagation favored by hydrogen at T=500K. Bonds are visualized (black lines) connecting iron atoms (gray). Hydrogen atoms are in gold.

\printbibliography[title={Supplementary References}]

\end{refsection}

\end{document}